%% file: manuscript.tex
\documentclass[acmsmall]{acmart}

\AtBeginDocument{%
  \providecommand\BibTeX{{%
    \normalfont B\kern-0.5em{\scshape i\kern-0.25em b}\kern-0.8em\TeX}}}

\setcopyright{acmcopyright}
\copyrightyear{2021}
\acmYear{2021}
\acmDOI{10.1145/XXXXX}

\begin{document}

\title{Metrics and Design of an Instruction Roofline Model for AMD GPUs}


\author{Matthew Leinhauser}
\email{mattl@udel.edu}
\orcid{0000-0003-2914-1483}
\affiliation{%
  \institution{Center for Advanced Systems Understanding}
  \city{Görlitz}
  \country{Germany}}
\affiliation{%
  \institution{University of Delaware}
  \city{Newark}
  \state{Delaware}
  \country{USA}}

\author{Ren{\'e} Widera}
\affiliation{%
  \institution{Helmholtz-Zentrum Dresden-Rossendorf Laboratory}
  \city{Dresden}
  \country{Germany}}
\email{r.widera@hzdr.de}

\author{Sergei Bastrakov}
\affiliation{%
  \institution{Helmholtz-Zentrum Dresden-Rossendorf Laboratory}
  \city{Dresden}
  \country{Germany}}
\email{s.bastrakov@hzdr.de}

\author{Alexander Debus}
\affiliation{%
  \institution{Helmholtz-Zentrum Dresden-Rossendorf Laboratory}
  \city{Dresden}
  \country{Germany}}
\email{a.debus@hzdr.de}

\author{Michael Bussmann}
\affiliation{%
  \institution{Center for Advanced Systems Understanding}
  \city{Görlitz}
  \country{Germany}}
\affiliation{%
  \institution{Helmholtz-Zentrum Dresden-Rossendorf Laboratory}
  \city{Dresden}
  \country{Germany}}
\email{m.bussmann@hzdr.de}

\author{Sunita Chandrasekaran}
\affiliation{%
  \institution{University of Delaware}
  \city{Newark}
  \state{Delaware}
  \country{USA}}
\email{schandra@udel.edu}

\renewcommand{\shortauthors}{Leinhauser, et al.}

\input{tex/abstract}



\maketitle

\input{tex/introduction}

\input{tex/relatedwork}

\input{tex/experimentalsetup}

\input{tex/amdroofline}

\input{tex/picongpu}

\input{tex/profilingtoolsused}

\input{tex/rooflinemodelcomparison}

\input{tex/conclusionfuture}







\input{tex/acknowledgement}

\bibliographystyle{ACM-Reference-Format}
\bibliography{references}

\end{document}

%% file: tex/abstract.tex
\begin{abstract}
Due to the recent announcement of the Frontier supercomputer, many scientific application developers are working to make their applications compatible with AMD (CPU-GPU) architectures, which means moving away from the traditional CPU and NVIDIA-GPU systems. Due to the current limitations of profiling tools for AMD GPUs, this shift leaves a void in how to measure application performance on AMD GPUs. In this paper, we design an instruction roofline model for AMD GPUs using AMD’s ROCProfiler and a benchmarking tool, BabelStream (the HIP implementation), as a way to measure an application’s performance in instructions and memory transactions on new AMD hardware. Specifically, we create instruction roofline models for a case study scientific application, PIConGPU, an open source particle-in-cell (PIC) simulations application used for plasma and laser-plasma physics on the NVIDIA V100, AMD Radeon Instinct MI60, and AMD Instinct MI100 GPUs. When looking at the performance of multiple kernels of interest in PIConGPU we find that although the AMD MI100 GPU achieves a similar, or better, execution time compared to the NVIDIA V100 GPU, profiling tool differences make comparing performance of these two architectures hard. When looking at execution time, GIPS, and instruction intensity, the AMD MI60 achieves the worst performance out of the three GPUs used in this work.
\end{abstract}

%% file: tex/introduction.tex
\section{Introduction}
One of the most invaluable resources for performance and analytical modeling of high performance computing (HPC) codes is the roofline model developed by Williams et al. in 2009~\cite{williams2009}. The roofline model offers a simple way to visually understand an application’s performance (in FLOPS) and its bottlenecks. Understanding performance and bottlenecks helps in optimizing codes. Thus, the roofline model has proved itself as an invaluable resource for performance and analytical modeling of HPC codes. Using the roofline model, application developers easily know what to optimize within a code, although it might be challenging to implement those optimizations. However, the traditional roofline model only gives specific optimization insights into a code. Thus, researchers have developed several extensions, with perhaps the most useful being the cache-aware roofline model (CARM)~\cite{ilic:CARM} and the hierarchical roofline model~\cite{yang2020hierarchical, williams2010roofline}. 
Both the CARM and hierachical roofline model extend the traditional roofline model beyond the traditional DRAM/HBM measurement to include the cache memories. Typically, only the L1 and L2 caches are included. To distinguish the CARM and hierarchical roofline model from other types of roofline models, in this work, we refer to them as the roofline performance model (RPM).

Another extension to the traditional roofline model, the instruction roofline model (IRM), was developed in 2019~\cite{ding2019instruction}. This model offers additional performance insights for an application beyond the RPM such as access patterns and instruction throughput. Creating an IRM is very similar to constructing an RPM. Instead of calculating maximum achieved GFLOPs for the compute ceiling, the maximum achieved billions of instructions per second (GIPS) is calculated. For the memory bandwidth ceiling, instead of using the measured bandwidth in GB/s, the GB/s bandwidth is divided by the size of a transaction (32 bytes) to use billions of transactions per second (GTXN/s). For clarity, GIPS is preferred over GFLOPs because instruction-level performance is important for the IRM. Similarly, GTXN/s is favored over GB/s because, as Ding and Williams describe, an execution-level load, such as a warp-load, can create up to 32 transactions on NVIDIA GPUs depending on the memory patterns used. Therefore, making the transaction the preferred memory unit for analyzing memory access.

Given that most of the leading supercomputers in the world currently contain NVIDIA GPUs~\cite{top500}, porting an application to work with AMD GPUs is an important open challenge. Due to the architecture and terminology differences between NVIDIA and AMD GPUs, the metrics traditionally gathered by hardware profilers are not applicable to use to create roofline models for AMD GPUs. 
In this work, we design an IRM for AMD GPUs using metrics from the ROCProfiler~\cite{rocProf}, AMD's hardware profiler, translate equations meant for NVIDIA GPUs to AMD GPU components, and create formulas specifically for AMD GPUs. PIConGPU, the application we look at in this work, was selected as one of the eight teams to take part in the Department of Energy's (DOE) Frontier Center for Accelerated Application Readiness (CAAR) effort~\cite{OLCF:CAAR}. Frontier, the Oak Ridge Leadership Computing Facility's (OLCF) newest supercomputer, will reach completion and start running programs in early 2022, and will contain AMD CPUs and AMD GPUs in each node.

This paper makes the following contributions:
\begin{itemize}
    \item Defines metrics and formulas needed to create an instruction roofline for state-of-the-art AMD GPUs. These metrics and formulas can provide a significant contribution to the design of AMD's profiling tool
    \item Provides a framework to model instruction-level performance of an HPC application on AMD GPUs
    \item Compares the performance of a plasma physics code, PIConGPU, using instruction roofline models. Finding out which architecture PIConGPU performs better on is unclear due to profiling tool limitations for AMD GPUs
\end{itemize}

%% file: tex/relatedwork.tex
\section{Related Work}\label{sec:relatedwork}
Of the existing research on instruction roofline models, the most similar to our work comes from Ding et al.~\cite{ding2019instruction}. They defined the formulas necessary to create an IRM and to plot achieved performance. The most significant difference from our work is that Ding et al.'s model focuses on creating instruction roofline models for NVIDIA GPUs. Specifically, some of the formulas are tied to NVIDIA's basic level of execution, the warp (32 threads). Since AMD's basic level of execution is the wavefront, all of the formulas defined in their work that use warps are unusable for AMD GPUs. It is worth noting that a wavefront can constitute 32 threads in some consumer GPUs (AMD RDNA2 GPUs), but this paper focuses on AMD's HPC GPUs, where the size of a wavefront is 64 threads. Additionally, the metrics they gathered to create the roofline model come from NVIDIA's legacy profiler, nvprof. This profiling tool is compatible only with NVIDIA GPUs, thus leaving a gap on which metrics to use for AMD GPUs.

Recently, there have been a few efforts to create an IRM for AMD GPUs. Richards et al.~\cite{richards} and Mehta et al.~\cite{mehtaevaluating} created instruction roofline models for the AMD Radeon Instinct MI60 GPU using proxy applications. Unfortunately, both papers lack sufficient details on how the roofline models were created. While some of the metrics used are explicitly mentioned, we do not know exactly how they were used to measure theoretical and achieved instruction-level performance. We build on these works by clearly defining which metrics are used to create an IRM and how they are used to plot achieved performance. Additionally, our work leverages AMD's state-of-the-art GPU, the AMD Instinct MI100. To the best of our knowledge, no other research uses the MI100 GPU (released in November 2020), or details formulas to create an IRM for AMD GPUs.

%% file: tex/experimentalsetup.tex
\section{Experimental Setup}
In this section, we go over the specifications of the machines we ran the PIConGPU simulations on, Summit and an early access Frontier Center of Excellence machine. We also discuss the specifications of the next-generation supercomputer, Frontier, that will run PIConGPU once built.

\subsection{Summit}
The Summit supercomputer, built from IBM AC922 nodes, currently sits in the OLCF at Oak Ridge National Laboratory in Oak Ridge, Tennessee, USA. Summit, at the time of writing, is the world's second fastest supercomputer \cite{top500}. The machine features 4608 nodes. Each node consists of two IBM Power9 CPUs (9,216 in total) and 6 NVIDIA Tesla V100 GPUs (27,648 in total). Each CPU contains 512 GB DDR4 RAM. In total, each node has 96 GB of High Bandwidth Memory (HBM) and uses NVIDIA's high-speed NVLink to communicate. 256 cabinets are used to contain the machine. The machine can achieve 200 peta-floating-point operations per second (PFLOPs), or $200 \times 10 ^{15}$ floating-point operations per second. The NVIDIA V100 GPUs use CUDA, which is a parallel computing platform and a programming model used exclusively for GPU computing on NVIDIA devices. On Summit we utilize Alpaka 0.6.0, CUDA 10.1.243, cupla 0.3.0-dev, Nsight Compute 2020.1, and Nsight Systems 2020.5.1.85.

\subsection{Early Access Frontier COE Machine}
The Early Access Frontier COE Machine (EAFCOEM) features eight nodes. Each node contains one high performance computing (HPC) and artificial intelligence optimized AMD EPYC CPU. In addition to the CPU, two nodes contain four NVIDIA Tesla V100 GPUs and the other six nodes contain four purpose built AMD Radeon Instinct GPUs. Of these six nodes, some contain Radeon Instinct MI60 GPUs while others hold AMD Instinct MI100 GPUs. As the release of Frontier nears, the nodes on this system will change to reflect the CPU and GPU devices the production Frontier machine will contain. Currently, the EAFCOEM supports both CUDA and HIP due to having both NVIDIA and AMD GPU devices. On the EAFCOEM, we utilize up to AMD ROCm 4.1.1. and HIP, Alpaka 0.6.0, and the latest stable commits of rocProf (up to and including commit 759f081) and BabelStream (up to and including commit 5182342).

\subsection{Frontier (2021-2022)}
Similar to the EAFCOEM, Frontier will hold one HPC and artificial intelligence optimized AMD EPYC CPU and four purpose built AMD Radeon Instinct GPUs per node. At the time of writing, we know Frontier will have over 100 cabinets to fit the nodes and use AMD Infinity Fabric to communicate. Frontier will use HIP, which is an open-source parallel computing model (an extension of C++) that can be used across multiple devices regardless of vendor. The peak performance is supposed to reach greater than 1.5 exa-FLOPs (EFLOPs), which is $1.5 \times 10^{18}$ floating point operations per second.

%% file: tex/amdroofline.tex
\section{Constructing Instruction Roofline Models for AMD GPUs}
\label{sec:construction}
In this section, we introduce AMD's ROCProfiler and explain which metrics are needed from it to measure an application's performance using an IRM. We then introduce the formulas used to create the IRM and show how the metrics gathered from the ROCProfiler can be applied to the IRM.

\subsection{Gathering Metrics Using the ROCProfiler}
The AMD ROCProfiler (rocProf) is a command line profiling analysis tool. This tool allows the user to get performance counters -- and derived metrics from those counters -- for an application. 
The tool works solely for applications using the ROCm accelerator backend. 
The metrics used for deriving IRMs on the NVIDIA V100 cannot be used on the AMD Radeon Instinct MI60 or AMD Instinct MI100 GPUs. The reason for this is because there is no way to extract the number of transactions from the L1 cache, L2 cache, or the DRAM/HBM using rocProf.
Instead, we use rocProf to get the FETCH\_SIZE, WRITE\_SIZE, SQ\_INSTS\_SALU, and SQ\_INSTS\_VALU metrics, and the kernel runtimes to construct a roofline model. 
The FETCH\_SIZE metric returns the total number of kilobytes (KBs) fetched from the GPU memory. 
Similarly, the WRITE\_SIZE metric returns the total number of KBs written to the GPU memory. 
Before using these metrics, we convert each value from KBs to bytes. 
The SQ\_INSTS\_SALU metric tells how many scalar-ALU instructions are issued to the GPU.
Similarly, the SQ\_INSTS\_VALU metric tells how many vector-ALU instructions are issued to the GPU.

\subsection{Using Metrics to Create an Instruction Roofline Model}
The IRMs presented here for the AMD MI60 and MI100 GPUs are built off of the work of Richards et al.~\cite{richards} from 2020. Instead of re-scaling the memory bandwidth to billions of transactions per second (GTXN/s), we leave the memory bandwidth in GB/s. 
Additionally, since there is no way to extract the number of transactions from rocProf, we use instructions per byte as the measurement unit on the horizontal axis,instead of instructions per transaction.

\begin{figure}[!t]
    \centering
    \includegraphics[scale=0.4]{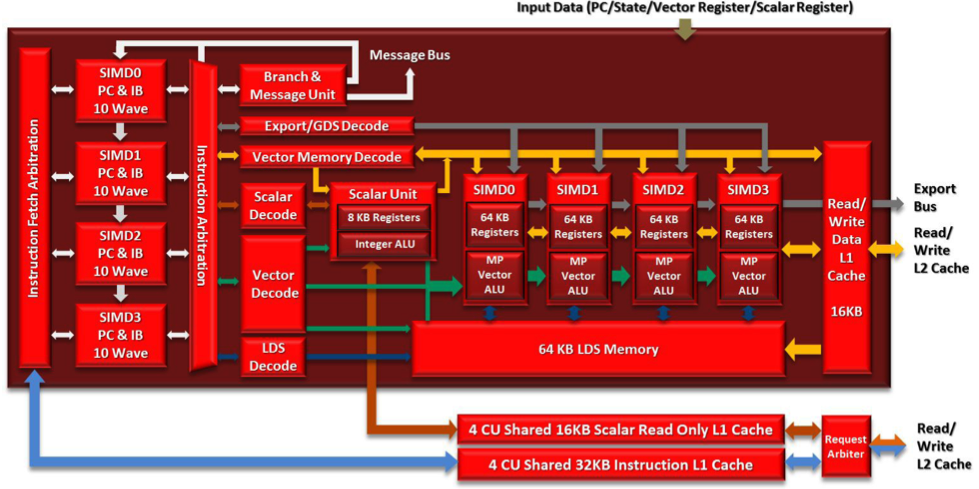}
    \caption{The compute unit for all AMD Graphics Core Next GPUs, which includes the MI60. The AMD CDNA GPUs (MI100) Compute Units are based off of the compute unit shown here. This image comes from the AMD Graphics Core Next white paper~\cite{devices2012amd}.}
    \label{fig:MI60_CU}
\end{figure}

To calculate the achieved GIPS and instruction intensity performance, we need to find the number of instructions issued. We show how to get the number of instructions in Equation~\ref{instructions_formula}. We multiply the SQ\_INSTS\_VALU metric by four because this metric gives the number of instructions issued per SIMD. For the AMD MI60 and MI100 GPUs, there are four SIMD vector units per compute unit. Figure \ref{fig:MI60_CU}, which comes from the original AMD Graphics Core Next (GCN) GPU white paper \cite{devices2012amd}, shows the four vector ALU units per SIMD. Similarly, we do not multiply the SQ\_INSTS\_SALU metric by anything because there is only one scalar unit per compute unit.

\begin{equation}\label{instructions_formula}
    instructions = (SQ\_INSTS\_VALU \times 4) + SQ\_INSTS\_SALU
\end{equation}

To calculate the instruction intensity performance, measured in instructions per byte, the FETCH\_SIZE and WRITE\_SIZE metrics from rocProf are used. The sum of those metrics is then multiplied by the kernel runtime and that quantity divides the number of wavefront scaled instructions. This is shown in Equation~\ref{II_AMD}. It is worth noting that because we normalize the instructions to the wavefront-level, we are actually calculating wavefront-level instruction intensity performance rather than a universal instruction intensity performance. Additionally, we calculate the the instruction intensity performance rather than the instruction intensity because of the limited metrics available to use with rocProf. Instruction intensity can be easily determined by calculating the throughput of GPU HBM reads and writes.

\begin{equation}\label{II_AMD}
    Instruction\ Intensity\ Performance = \frac{\frac{instructions}{64}}{(bytes\ read + bytes\ written) \times runtime}
    \vspace{5mm}
\end{equation}

To calculate the peak theoretical GIPS, we modify the peak GIPS equation from~\cite{ding2019instruction} to work with AMD architecture. AMD uses the term~\textit{compute units} instead of~\textit{streaming multiprocessors}. The MI60 and MI100 contain 64 and 120 compute units (CU) respectively. Additionally, AMD GPUs use wavefronts instead of warps. The MI60 and MI100 GPUs each contain one wavefront scheduler per compute unit (WFS/CU). The theoretical instructions per cycle (IPC) variable is 1 (1 IPC) as stated in~\cite{amdhiptraining}. This is in unison with~\cite{ding2019instruction} despite focusing on a different GPU vendor. The frequency is measured in gigahertz as shown in Equation~\ref{GIPS_AMD}.

\begin{equation}\label{GIPS_AMD}
    GIPS_{peak} = CU \times WFS/CU \times IPC \times frequency
    \vspace{5mm}
\end{equation}

The achieved instruction performance (GIPS$_{achieved}$) in GIPS is calculated by the formula shown in Equation~\ref{achieved_GIPS_AMD}. 
We divide by 64 because 64 threads constitute a wavefront in the AMD GPUs we target. The number of instructions is calculated as shown in Equation~\ref{instructions_formula}

\begin{equation}\label{achieved_GIPS_AMD}
    GIPS_{achieved} = \frac{\frac{instructions}{64}}{1 \times 10^{9} \times runtime}
    \vspace{5mm}
\end{equation}

The IRMs for AMD GPUs could easily re-scale the bandwidth into GTXN/s as shown for the V100 IRMs, and this might seem like a more equal comparison, but since we cannot get the number of transactions to use for the instruction intensity/instruction intensity performance, we did not want to offer a misleading comparison.

%% file: tex/picongpu.tex
\section{PIConGPU, a Plasma Physics Application}
\label{sec:picongpu}
\begin{figure}[!h]
    \centering
    \includegraphics[scale=0.5]{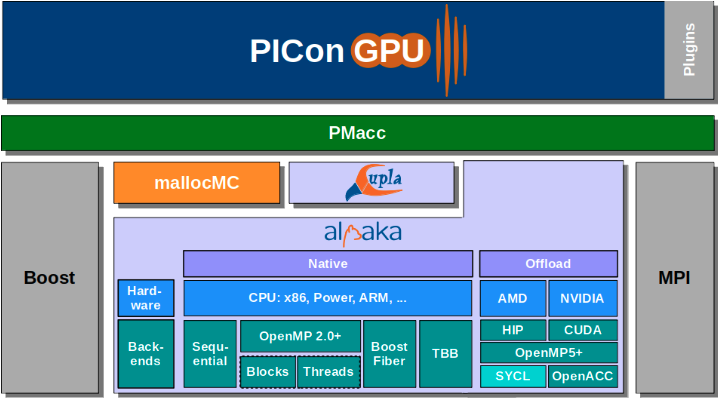}
    \caption{The software stack of PIConGPU. Due to the abstraction Alpaka brings, only a few top level changes were made to get PIConGPU running on AMD GPUs via HIP.}
    \label{fig:picSoftwareStack}
\end{figure}

PIConGPU~\cite{PIConGPU2013} is an open source particle-in-cell (PIC) simulations application that runs on general purpose GPUs for plasma and laser-plasma physics used to develop advanced particle accelerators for radiation therapy of cancer, high-energy physics, and photon science. PIConGPU utilizes the Alpaka~\cite{Alpaka} backend and the particle-in-cell algorithm for its science case simulations. Alpaka is an open-source abstraction library written in C++14 that aims to provide performance portability across accelerators through the abstraction of underlying levels of parallelism. It is platform independent and also supports concurrent and cooperative use between the host device and any attached accelerators. Alpaka is used on top of HIP and therefore, most of the porting is done in Alpaka rather than PIConGPU. Due to the software stack PIConGPU uses, only a few top level changes are made to support running on AMD GPUs via HIP. Figure \ref{fig:picSoftwareStack} shows the entire PIConGPU software stack. More information about porting PIConGPU can be found in the 2016 ICHPC paper~\cite{zenker2016performance} and the Alpaka code repository~\cite{alpakarepo}.
As part of the ongoing CAAR for Frontier effort, we analyze the performance of PIConGPU on OLCF's Summit supercomputer, the second fastest in the world~\cite{top500} as of the time of writing, and on an early access Frontier Center of Excellence machine. 

In this work, we run PIConGPU's Traveling Wave Electron Acceleration (TWEAC) and Laser Wakefield Acceleration (LWFA) simulations.
A deeper dive into the science of the TWEAC and LWFA simulations is explained by Debus et al.~\cite{TWEAC}. We profile the simulations using various state-of-the-art profiling tools and micro-kernel benchmarking suites. Using the metrics gathered from profiling, we construct roofline models for PIConGPU's MoveAndMark and ComputeCurrent kernels. These kernels are measured as the most computationally intensive~\cite{leinhauser2021performance} and are called in every PIConGPU simulation run. Therefore, optimizing these kernels will improve every simulation in PIConGPU. By constructing roofline models, we determine the future optimizations needed to exploit the best performance of PIConGPU.

%% file: tex/profilingtoolsused.tex
\section{Profiling Tools Used}
\label{sec:profile}
To profile the TWEAC and LWFA simulations on the NVIDIA V100 GPU, we use NVIDIA’s legacy profiling tool, NVProf~\cite{NVProf}, and its state-of-the-art profiling tools, Nsight Compute~\cite{NsightCompute} and Nsight Systems~\cite{Nsys}. These profiling tools can only target NVIDIA GPUs. Hence we cannot use these tools to profile simulations on AMD GPUs.

To profile PIConGPU on the AMD devices, we initially used rocProf.
However, we quickly found that rocProf did not suffice for acquiring all the metrics needed to construct IRMs and thus we pivoted to use various benchmarking suites. These benchmarking suites filled in most of the gaps needed to create roofline models for the AMD devices. As AMD architecture continues to grow in use, we hope to see new profiling tools with broader capabilities released in the future. 
The following subsections narrate findings using NVIDIA's tools and the micro-kernel benchmarking suites we used.

In making the roofline plots shown later in this report, we utilize the metrics the NERSC Roofline-on-NVIDIA-GPUs code repository uses~\cite{NERSC_code_repo}. The data gathered using rocProf, Nsight Compute, nvprof, and the HIP implementation of BabelStream can be found in a repository we created on GitHub~\cite{roofline_plot_code_repo}. Additionally, modifications we made to the NERSC code repository can be found in the same repository.

\subsection{Nsight Compute and Nsight Systems}
Nsight Compute is NVIDIA’s latest application profiling tool. It boasts similar functionality to NVProf, but does not map the application runtime by kernel calls (as NVProf did). One of the biggest enhancements, the Roofline Analysis feature, came to Nsight Compute in version 2020.1. The Roofline Analysis feature automatically generates a roofline plot within the profiling report that is output. This work utilizes the Roofline Analysis feature and expands upon its initial use by utilizing Nsight Compute to create custom roofline plots.

\begin{figure}[!t]
    \centering
    \includegraphics[scale=0.3]{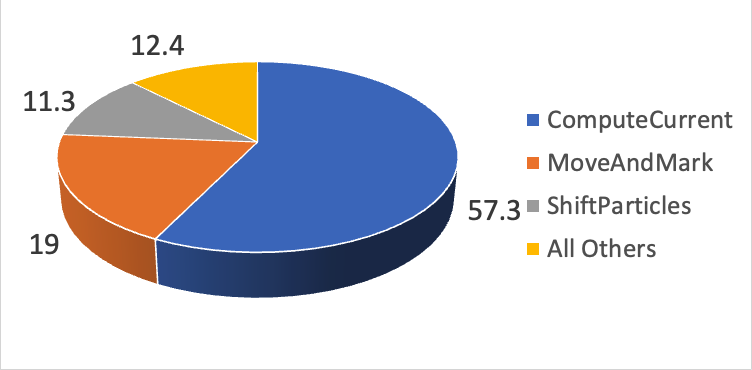}
    \caption{Execution time (\%) for different kernels within PIConGPU's TWEAC science case. The MoveAndMark and ComputeCurrent kernels take up over 75\% of the overall runtime.}
    \label{fig:TWEAC_Pie_Chart}
\end{figure}

Nsight Systems visually maps an application from execution to termination. The visualization, that is in the form of a timeline, is useful for deducing which kernels take up the most execution time, which bottlenecks exist in the code, and which kernels under-perform. This work utilized Nsight Systems to verify the developers’ rationale about PIConGPU’s most computationally-intensive kernels and to find out the percentage of runtime those kernels took up. Figure ~\ref{fig:TWEAC_Pie_Chart} shows a visualization of the timeline from Nsight Systems to focus on our kernels of interest.

\subsection{Micro-kernel Benchmark Tools}
To gather the memory bandwidth for the AMD MI60 and MI100 devices, we use a variety of micro-kernel benchmarking tools. 
First, we use the gpumembench~\cite{gpumembench} Benchmark Suite as a way to assess on-chip GPU memory bandwidth.
Using the programs in the suite, we measure the instruction throughput, shared memory operations, and constant memory operations on the MI60 and MI100 GPUs. 
The other benchmark tool we use is BabelStream~\cite{BabelStream}. 
Formerly called GPU-Stream, BabelStream measures memory transfer rates to and from the global device memory on GPUs. 
BabelStream differs from other GPU memory bandwidth benchmarks and benchmarking suites in that it does not include PCIe transfer time in its results. 
BabelStream provides memory bandwidth results that are attainable.
The output for the copy functions 808,975.476 MB/s for the MI60 GPU and 933,355.781 MB/s for the MI100 GPU) from BabelStream is used to represent the memory bandwidth for the AMD MI60 and MI100 GPU IRMs. It is worth noting that the measured bandwidth can vary when using an implementation of BabelStream other than the HIP implementation (for example the AOMP implementation).
On the roofline plots we convert each measurement to GB/s.

%% file: tex/rooflinemodelcomparison.tex
\section{Roofline Model Comparisons}
\label{sec:roofline}
\begin{figure}[!t]
    \centering
    \includegraphics[scale=0.32]{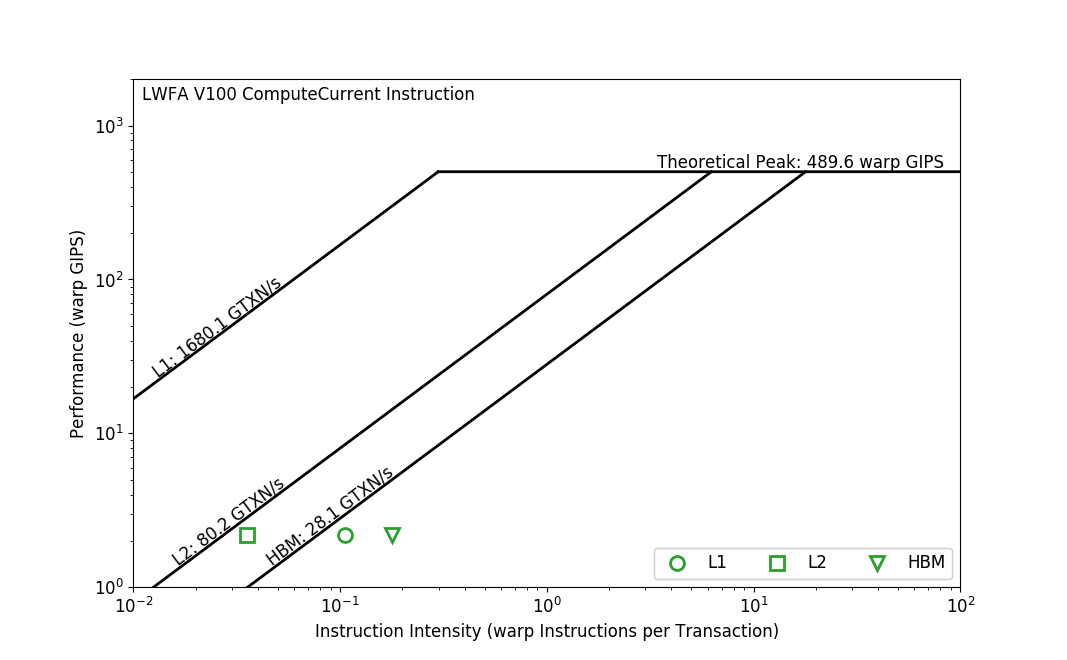}
    \caption{The instruction roofline model for the ComputeCurrent kernel in the LWFA simulation on the \textbf{NVIDIA V100 GPU} on OLCF's Summit. Looking at the plotted points, one can see this kernel's instruction-level performance can be greatly improved.}
    \label{fig:LWFA_ComputeCurrent_V100_Instruction_Roofline}
\end{figure}

In this section we show IRMs generated on the NVIDIA V100, AMD MI60, and AMD MI100 for the PIConGPU LWFA science case, as seen in Figures~\ref{fig:LWFA_ComputeCurrent_V100_Instruction_Roofline}, \ref{fig:LWFA_V100_IRM_IIPB} and~\ref{fig:LWFA_ComputeCurrent_AMD_Instruction_Roofline}, and compare the results in table~\ref{tab:LWFA_ComputeCurrent_Comparison}. Additionally we show an IRM for the AMD MI60 and AMD MI100 for PIConGPU's TWEAC science case (Figure~\ref{fig:TWEAC_ComputeCurrent_AMD_Instruction_Roofline}). Table~\ref{tab:TWEAC_ComputeCurrent_Comparison} shows the same metrics that Table~\ref{tab:LWFA_ComputeCurrent_Comparison} did but for the TWEAC simulation. The IRM plots were created by modifying the scripts from the NERSC Roofline-On-NVIDIA-GPUs code repository. Finally, we dedicate a subsection to discussing the differences between AMD and NVIDIA GPU hardware and profiling tools to explain some of the differences outlined in this section.

\subsection{NVIDIA GPU Instruction Roofline Models}
\begin{figure}
    \centering
    \includegraphics[scale=0.32]{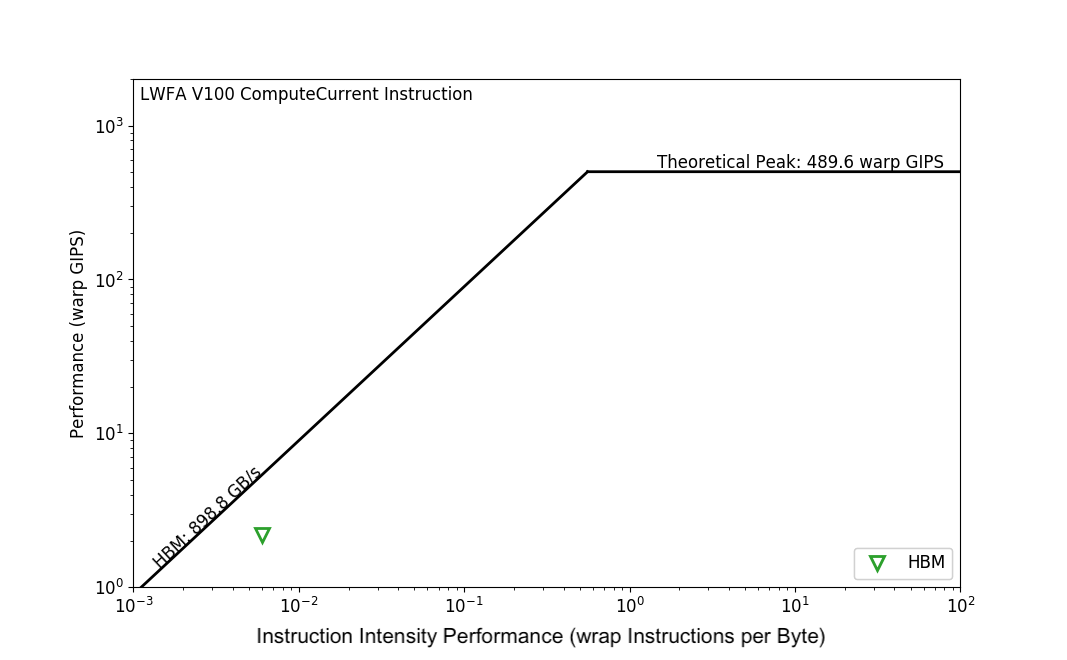}
    \caption{The instruction roofline model for the ComputeCurrent kernel in the LWFA simulation on the \textbf{NVIDIA V100 GPU} on OLCF's Summit. Here, the instruction intensity is measured in instructions/byte rather than instructions/transaction. This plot shows that there is much room for improvement.}
    \label{fig:LWFA_V100_IRM_IIPB}
\end{figure}

To construct the roofline models on the NVIDIA V100, we collected the same metrics from NVProf as mentioned in~\cite{ding2019instruction}.
The roofline model shown in Figure~\ref{fig:LWFA_ComputeCurrent_V100_Instruction_Roofline} represents the IRM for an instance of the ComputeCurrent kernel ran during an LWFA science case simulation of PIConGPU. 
The plot shows the compute ceiling measured in GIPS and the memory bandwidth ceiling in GTXN/s, along with the achieved warp GIPS and instruction intensity for the HBM, and the L1 and L2 caches. We observe in~\cite{ding2019instruction} that L1 points appearing on the left side of the plot (i.e. L1 points that have low instruction intensity) are more likely to show that the kernel has strided memory access patterns. We confirmed this is true for Figure~\ref{fig:LWFA_ComputeCurrent_V100_Instruction_Roofline} by following the method outlined in~\cite{ding2019instruction} under "Global Memory Walls".
Similarly, L2 points appearing more left on the plot represent 32-way bank conflicts. Using Nsight Compute, we confirm this kernel experiences many bank conflicts. We also see that it is HBM-bound and its computational performance cannot increase  without addressing the memory issues present. A difference between the plot in Figure~\ref{fig:LWFA_ComputeCurrent_V100_Instruction_Roofline} and the IRMs shown in~\cite{ding2019instruction} is that the V100 GPU utilized to run the PIConGPU LWFA simulation achieved a greater memory bandwidth than the one Ding et al. used thus leading to higher GTXN/s for the L1 cache, L2 cache, and HBM. Additionally, we also created an IRM for this kernel measuring instruction intensity in \textit{instructions/byte} to give a better comparison between NVIDIA and AMD. Figure~\ref{fig:LWFA_V100_IRM_IIPB} shows the instruction intensity measured in instructions/byte. We avoid including the L1 and L2 caches so that it is more similar to its AMD counterparts.

\subsection{AMD GPU Instruction Roofline Models}
\begin{figure}[!t]
    \centering
    \includegraphics[scale=0.32]{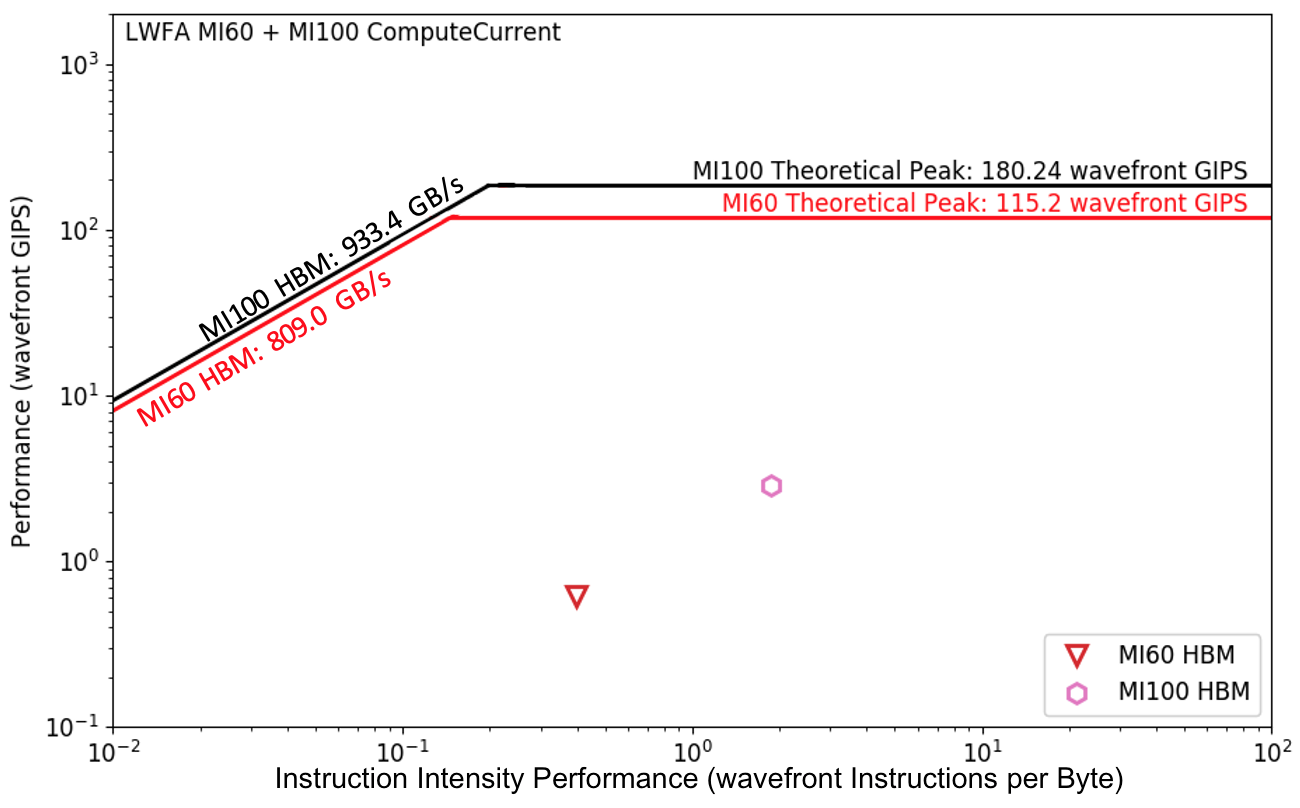}
    \caption{The instruction roofline model for the ComputeCurrent kernel in the LWFA simulation on the \textbf{AMD MI60 and MI100 GPUs}. This simulation was run on an AMD/Cray early-access system at OLCF. The HBM point in this roofline plot appears in a much better position than than the HBM point for the V100's roofline plot.}
    \label{fig:LWFA_ComputeCurrent_AMD_Instruction_Roofline}
\end{figure}

We follow the process of constructing IRMs outlined in Section~\ref{sec:construction} and find that the AMD MI60 GPU has a theoretical peak GIPS of 115.2 while the MI100 boasts a theoretical peak GIPS of 180.24. The memory bandwidth was measured using BabelStream. Due to AMD profiling tool limitations, we are unable to measure performance for the L1 and L2 caches on any AMD GPU. Figure~\ref{fig:LWFA_ComputeCurrent_AMD_Instruction_Roofline} shows the IRMs for both the AMD MI60 and MI100 on the LWFA science case's ComputeCurrent kernel. Even though we can only measure the HBM performance, we can assume that the L1 and L2 cache performance points would appear to the left of the HBM performance points. Following this assumption will give us synonymous information as the V100's IRM: the kernel has many bank conflicts and strided memory access. We want to stress the fact that we are \textit{assuming} the AMD IRMs are showing bank conflicts and strided memory access. We are not able to confirm these assumptions because of the limited amount of performance counter metrics available in AMD profiling tools.

Table 1 shows the achieved GIPS performance, Instruction Intensity, and the runtime of the ComputeCurrent kernel on the NVIDIA V100, AMD MI60 and MI100 GPUs for the LWFA science case. The table shows the kernel executed significantly faster on the V100 and MI100 than the MI60, with the MI100 having the best execution time overall.  The instruction intensities in the table are measured in $instructions/bytes$. According to~\cite{ding2019instruction}, the instruction intensity should be measured in $instructions/transactions$. Due to performance counter limitations on rocProf, we could not extract the amount of transactions. On the NVIDIA V100, the instruction intensity was measured as 0.178 instructions per transactions (as shown in Figure~\ref{fig:LWFA_ComputeCurrent_V100_Instruction_Roofline}). To make an equal comparison the V100's instruction intensity is measured in instructions per byte in Table ~\ref{tab:LWFA_ComputeCurrent_Comparison}.

\begin{table}[ht!]
    \centering
    \caption{Execution time, achieved GIPS, and instruction intensity for the LWFA simulation's ComputeCurrent kernel on the NVIDIA V100 \& AMD MI60 and MI100 GPUs. The values in this table are rounded to three decimal points and therefore manually calculating Achieved GIPS and Instruction Intensity may vary slightly from the numbers shown here. For execution time, a lower number is better. For GIPS and Instruction Intensity, a higher number is better.}
    \resizebox{14cm}{!}
    {
        \begin{tabular}{|c | c | c | c|}
            \hline
            \textbf{PIConGPU LWFA Simulation} & \multicolumn{3}{c|}{\textbf{ComputeCurrent}} \\
            \hline
            \textbf{GPU} & V100 & MI60 & MI100 \\
            \hline
            \textbf{Execution Time (s)} & 0.0040 & 0.0127 & 0.0025 \\
            \hline
            \textbf{\{Compute Units, Streaming Multiprocessors\}} & 80 & 64 & 120 \\
            \hline
            \textbf{Instructions/Cycle} & 1 & 1 & 1 \\
            \hline
            \textbf{Frequency (GHz)} & 1.530 & 1.800 & 1.502 \\
            \hline
            \textbf{\{Wavefront, Warp\} Schedulers} & 4 & 1 & 1 \\
            \hline
            \textbf{Peak GIPS} & 489.60 & 115.20 & 180.24 \\
            \hline
            \textbf{Achieved GIPS} & 2.178 & 0.620 & 2.856 \\
            \hline
            \textbf{Instructions} & 279,498,240 & 502,440,960 & 449,796,480 \\
            \hline
            \textbf{Bytes Read} & 267,280,000,000 & 1,125,436,000 & 1,124,711,000 \\
            \hline
            \textbf{Bytes Written} & 97,329,000,000 & 432,711,000 & 408,483,000 \\
            \hline
            \textbf{\{Wavefront, Warp\}-Level Instruction Intensity (inst/byte)} & 0.006 & 0.398 & 1.863 \\
            \hline
        \end{tabular}
    }
    \label{tab:LWFA_ComputeCurrent_Comparison}
\end{table}

\begin{figure}[!t]
    \centering
    \includegraphics[scale=0.32]{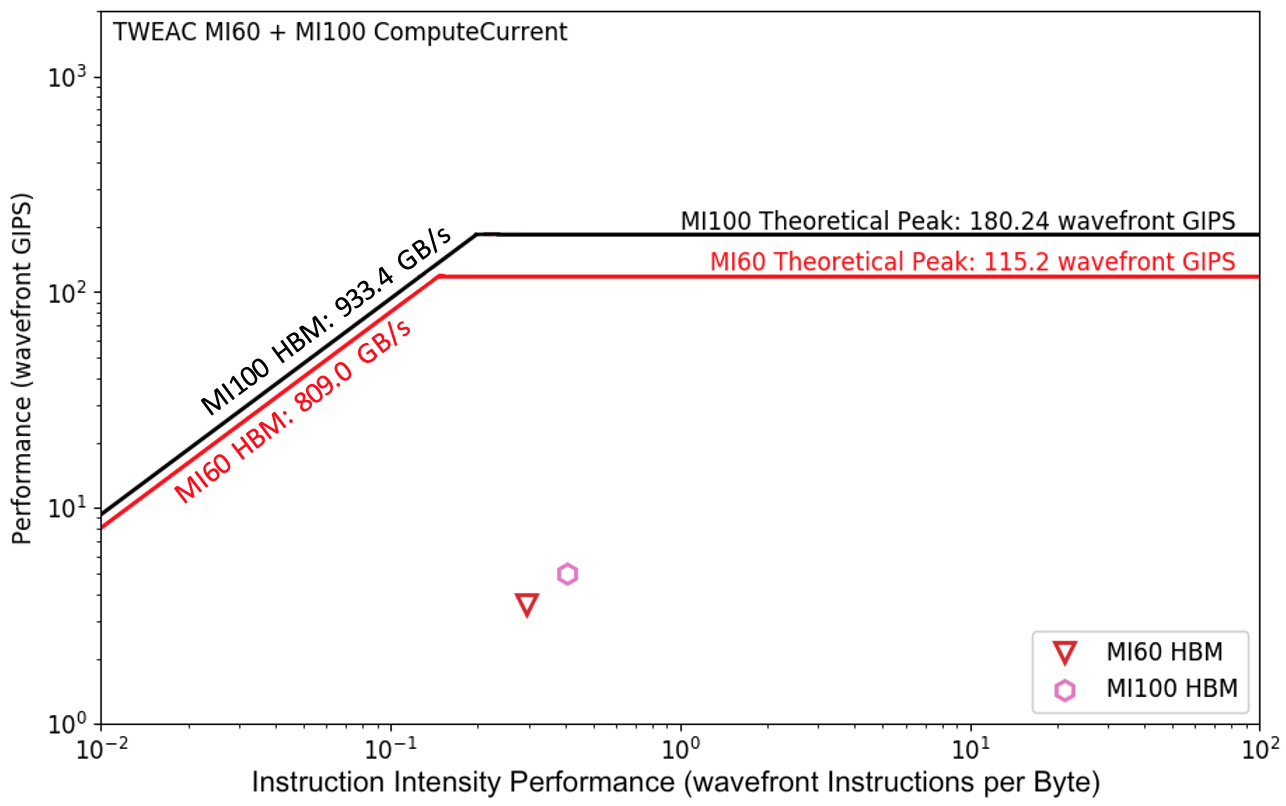}
    \caption{The instruction roofline model for the ComputeCurrent kernel in the TWEAC simulation on the \textbf{AMD MI60 and MI100 GPUs}. This simulation was run on an AMD/Cray early-access system at OLCF. Unlike its NVIDIA V100 counterpart, this roofline plot offers no insight into the L1 and L2 cache performance due to a lack of performance counters on the AMD GPUs.}
    \label{fig:TWEAC_ComputeCurrent_AMD_Instruction_Roofline}
\end{figure}

Looking at the ComputeCurrent kernel for PIConGPU's TWEAC simulation we see similar results to its LWFA simulation for the same kernel. Table \ref{tab:TWEAC_ComputeCurrent_Comparison} shows that the MI100 still has the fastest execution time of the kernel, but its achieved GIPS is much lower than the V100. To measure instructions on AMD GPUs, we use the vector-ALU and scalar-ALU instructions stated by rocProf. The V100 uses an nvprof metric ($inst\_executed$) that measures all the different types of instructions the profiled kernel uses. Just as in the table above, the instruction intensity for the V100 in Table \ref{tab:TWEAC_ComputeCurrent_Comparison} is reported in instructions per byte to make an equal comparison with AMD GPUs. The instruction intensity on the V100 in instructions per transaction is 4.931. In Tables \ref{tab:LWFA_ComputeCurrent_Comparison} and \ref{tab:TWEAC_ComputeCurrent_Comparison}, the NVIDIA V100 processes a much greater number of instructions and reads and writes a significantly higher number of bytes to the DRAM than the AMD GPUs. At the time of writing, we are unsure of why this behavior is occurring and continue to look into it. Figure~\ref{fig:TWEAC_ComputeCurrent_AMD_Instruction_Roofline} shows the instruction roofline model for PIConGPU's TWEAC simulation ComputeCurrent kernel. 

\begin{table}[ht!]
    \centering
    \caption{Execution time, achieved GIPS, and instruction intensity for the TWEAC simulation's ComputeCurrent kernel on the NVIDIA V100 \& AMD MI60 and MI100 GPUs. The values in this table are rounded to three decimal points and therefore manually calculating Achieved GIPS and Instruction Intensity may vary slightly from the numbers shown here. For execution time, a lower number is better. For GIPS and Instruction Intensity, a higher number is better.}
    \resizebox{14cm}{!}
    {
        \begin{tabular}{|c | c | c | c|}
            \hline
            \textbf{PIConGPU TWEAC Simulation} & \multicolumn{3}{c|}{\textbf{ComputeCurrent}} \\
            \hline
            \textbf{GPU} & V100 & MI60 & MI100 \\
            \hline
            \textbf{Execution Time (s)} & 0.283 & 0.394 & 0.246 \\
            \hline
            \textbf{\{Compute Units, Streaming Multiprocessors\}} & 80 & 64 & 120 \\
            \hline
            \textbf{Instructions/Cycle} & 1 & 1 & 1 \\
            \hline
            \textbf{Frequency (GHz)} & 1.530 & 1.800 & 1.502 \\
            \hline
            \textbf{\{Wavefront, Warp\} Schedulers} & 4 & 1 & 1 \\
            \hline
            \textbf{Peak GIPS} & 489.60 & 115.20 & 180.24 \\
            \hline
            \textbf{Achieved GIPS} & 6.634 & 3.586 & 4.993 \\
            \hline
            \textbf{Instructions} & 60,149,000,000 & 90,319,028,127 & 78,488,570,820  \\
            \hline
            \textbf{Bytes Read} & 40,931,000,000 & 11,451,009,000 & 11,460,394,000 \\
            \hline
            \textbf{Bytes Written} & 1,810,100,000 & 785,101,000  & 792,172,000 \\
            \hline
            \textbf{\{Wavefront, Warp\}-Level Instruction Intensity (inst/byte)} & 0.155 & 0.293 & 0.408 \\
            \hline
        \end{tabular}
    }
    \label{tab:TWEAC_ComputeCurrent_Comparison}
\end{table}

\subsection{Discussion of Rooflines, Profiling Tools, and GPU Hardwares}
As we alluded to earlier, there are many differences between AMD and NVIDIA GPUs. The first main difference is the size of a warp and a wavefront. One possible explanation for the difference between achieved GIPS on NVIDIA GPUs and AMD GPUs could be the size of a wavefront (64 threads) vs. the size of a warp (32 threads). Since a wavefront is twice as big as a warp, achieved GIPS on AMD GPUs are at a significant disadvantage when compared head to head with NVIDIA GPUs due to how instructions are scaled. If each GPU issued 100,000 compute instructions, the achieved GIPS for an NVIDIA GPU would be twice as high as the AMD GPU's solely due to the idea of scaling instructions to the warp-/wavefront-level. However, this is necessary in order to figure out the instruction-level performance of the GPU. Additionally, nvprof's $inst\_executed$ metric measures \textit{all} types of instructions issued by the NVIDIA GPU whereas the vector-ALU and scalar-ALU instructions comprise \textit{only} the compute instructions for AMD GPUs. We decided to utilize the $inst\_executed$ metric for the experiments conducted in this work because both nvprof and Nsight Compute do not offer a single metric to extract only the compute instructions issued to an NVIDIA GPU like rocProf does. This means that the NVIDIA GPU might contain instructions that have nothing to do with the compute instructions of the kernel being measured as opposed to the instructions metrics we collect for AMD GPUs, which only contain compute instructions. However, Nsight Compute does allow users to obtain detailed per-instruction-type metrics which can then be aggregated across all compute instructions.

Looking beyond what constitutes the achieved points and instead focusing on the base rooflines themselves, we also see a few differences. The first noted difference is the memory bandwidth for each GPU. Using Nsight Compute, we see the achieved bandwidth of the NVIDIA V100 GPU (shown in Figure~\ref{fig:LWFA_V100_IRM_IIPB} is over 99\% of its theoretical bandwidth (900 GB/s). With rocProf, we cannot measure the achieved bandwidth for AMD GPUs so we use the HIP implementation of BabelStream to do so. Using BabelStream, the AMD MI60 achieves ~81\% of its theoretical bandwidth and the MI100 achieves ~78\%. The compute ceilings also show different stories. The theoretical GIPS ceiling for the V100 is about 2.7x higher than the MI100's and 4.25x higher than the MI60's. As Ding and Williams described in~\cite{ding2019instruction}, the theoretical GIPS ceiling is calculated using a GPU's hardware components. The biggest differences that lead to the lower theoretical GIPS ceiling for the AMD MI60 are the number of compute units (64 vs. the V100's 80 and MI100's 120) and the 1 wavefront scheduler per compute unit (vs. the V100's 4 warp schedulers per streaming multiprocessor). Despite having the highest frequency out of all three GPUs used, these two differences give the MI60 to have the lowest theoretical GIPS ceiling. Similar to the MI60, the MI100's GIPS ceiling is low compared to the V100's because the MI100 only has 1 wavefront scheduler per compute unit. The warp/wavefront schedulers per compute unit/streaming multiprocessor are what drive the theoretical GIPS up. For example, if the V100 only had 1 warp scheduler per streaming multiprocessor, its theoretical GIPS ceiling would be only 122.4, a quarter of what it is now. Due to these fundamental differences between GPU hardware and their profiling tools, it is hard to create a instruction-level performance model that will equally relate GPUs made by different vendors.

Finally, the last point we want to make is that while roofline models offer many insights into a kernel's performance, they do not explicitly show the performance limiters of that kernel. While showing memory bandwidth, instruction throughput or FLOPS under-utilization is helpful, the real performance limiters (memory stalls, instruction dependencies, etc.) must be analyzed separately. Adding on to the difficulty of optimizing kernels is that some performance limiters can be architecture dependent. In this paper, all three of the GPUs used have many similarities and therefore improvements and refactoring to reduce and/or avoid performance limiters have positive effects for all devices. Alpaka provides the infrastructure necessary to write unified code, but how memory is mapped to workers/threads depends on the architecture and how it is abstracted for PIConGPU using PMacc (see Figure~\ref{fig:picSoftwareStack}).

%% file: tex/conclusionfuture.tex
\section{Conclusions and Future Work}
\label{sec:conclude_future}
In this paper, we showed how to construct IRMs for AMD GPUs using metrics from rocProf and micro-kernel benchmarking suites.
We also highlighted the existing limitations that make measuring performance on AMD GPUs a challenge. 
Finally, we built and compared IRMs for PIConGPU's TWEAC and LWFA simulations. Specifically, we looked at the LWFA's ComputeCurrent kernel on different hardware (NVIDIA vs. AMD) and compared the performances of the hardware used.
We find that the result of which hardware PIConGPU performs better on is unclear because of the profiling tool limitations for AMD GPUs. Additionally, comparing the achieved instruction-level performance between AMD and NVIDIA is difficult due to the limitations of each vendor's profiling tool(s).

In the future, we hope to expand upon this work by designing and constructing roofline models, along with analyzing the performance of PIConGPU, on future AMD GPUs found in the Frontier supercomputer. We will continue to look into why the AMD MI100 is processing more instructions than the NVIDIA V100, and why the V100 is reading and writing more bytes to the DRAM than the AMD GPUs. Similarly, we wish to identify how many additional instructions are added due to the intrusion of profiling tools. Finally, we will investigate how to extract the achieved FLOPs and the number of L1 and L2 cache, and HBM/DRAM read and write transactions from AMD GPUs to allow for more equal comparisons between NVIDIA and AMD GPUs.

%% file: tex/acknowledgement.tex
\begin{acks}
The authors thank Ronnie Chatterjee, the former PIConGPU CAAR liaison for his invaluable support. Additionally, we extend gratitude to Nicholas Malaya, the AMD Technical Lead for the Frontier Center of Excellence for his help with understanding AMD GPUs and to Felix Schmitt of NVIDIA for his help on understanding Nsight Compute metrics. 
We extend thanks to Nan Ding \& Sam Williams 
at LBNL for their insights into instruction roofline models. Our thanks to Shirley Moore, of UTEP, and David Richards, of LLNL, for their help in understanding rocProf. Finally, our thanks to David Rogers of ORNL and our current CAAR liaison for the valuable discussions.

This work was partly funded by the Center for Advanced Systems Understanding (CASUS) which is financed by the German Federal Ministry of Education and Research (BMBF) and by the Saxon Ministry for Science, Art, and Tourism (SMWK) with tax funds on the basis of the budget approved by the Saxon State Parliament.

This research partially used resources of the Oak Ridge Leadership Computing Facility (OLCF) at the Oak Ridge National Laboratory, which is supported by the Office of Science of the U.S. Department of Energy under Contract No. DE-AC05-00OR22725.

This material is based upon work supported by the National Science Foundation (NSF) under grant no. 1814609.
\end{acks}